\documentclass{iopart}
\usepackage{graphicx}

\newcommand{\bra}[1]{\left\langle #1\right|}
\newcommand{\ket}[1]{\left| #1\right\rangle}
\newcommand{\braket}[2]{\left\langle #1|#2\right\rangle}
\newcommand{\ketbra}[2]{\left| #1\right\rangle\!\left\langle#2\right|}

\begin{document}

\title[Maximum likelihood in homodyne tomography]{Iterative maximum-likelihood reconstruction in quantum homodyne tomography}

\author{A. I. Lvovsky\dag\
\footnote{URL:
http://www.uni-konstanz.de/quantum-optics/quantech/} }

\address{\dag\ Fachbereich Physik, Universit\"at Konstanz, D-78457 Konstanz, Germany}

\begin{abstract}
I propose an iterative expectation maximization algorithm for
reconstructing the density matrix of an optical ensemble from a
set of balanced homodyne measurements. The algorithm applies
directly to the acquired data, bypassing the intermediate step of
calculating marginal distributions. The advantages of the new
method are made manifest by comparing it with the traditional
inverse Radon transformation technique.
\end{abstract}


\submitto{\JOB}

% \maketitle

%A quantum state of a system, however simple, cannot be determined
%by a single measurement. Neither will repeated, identical
%measurements performed on multiple identical copies of a quantum
%state generally will not yield a complete information about the
%ensemble in question. Such a set of measurements will however
%provide the probability distribution of the ensemble measured over
%the eigenstates of the measurement apparatus.

\emph{Quantum tomography} is a technique of characterizing a state
of a quantum system by subjecting it to a large number of quantum
measurements, each time preparing the system anew. By varying the
configuration of the measurement apparatus, one acquires the
quantum statistics associated with different bases from which
complete information about the state of the system can be
extracted.

The ensemble's density matrix can be evaluated from the
experimental statistical data by a number of techniques. In this
paper we are dealing with one such technique, the \emph{Maximum
Likelihood} (MaxLik) estimation. Assuming a particular density
matrix $\hat{\rho}$, one can evaluate the likelihood (probability)
of acquiring a particular set of measurement results. The ansatz
of the MaxLik method is to find, among the variety of all possible
density matrices, the one which maximizes the probability of
obtaining the given experimental data set. To date, this method
has been applied to various quantum and classical problems from
quantum phase estimation \cite{Qphase} to reconstruction of
entangled optical states \cite{Rehacek01,White01}.

In the present work we discuss applications of likelihood
maximization to \emph{quantum homodyne tomography} - a method of
characterizing a quantum state of an optical mode by means of
multiple phase-sensitive measurements of its electric field's
quantum noise. Since its proposal \cite{VogelRisken} and first
experimental implementation \cite{Raymer93} in the early 1990s,
quantum homodyne tomography has become a robust and versatile tool
of quantum optics and has been applied in many different
experimental settings. As any statistical method, it is compatible
with the likelihood maximization approach. Yet so far
experimentalists have used a clumsier and less accurate
reconstruction algorithms based on the inverse Radon
transformation. The goal of this paper is to provide an explicit
numerical recipe for the MaxLik reconstruction of a quantum state
from homodyne data and to demonstrate its advantages by applying
it to \emph{bona fide} experimental results.

The applications of MaxLik estimation to homodyne tomography have
been investigated by Banaszek, who has reconstructed the
photon-number distribution (the diagonal density matrix elements
which correspond to a phase-randomized optical ensemble) from a
Monte-Carlo simulated data set \cite{B1,B2}. In a subsequent
publication \cite{B5}, Banaszek {\it et al.} discussed the MaxLik
estimation of the complete density matrix, but no explicit
reconstruction algorithm has been presented.

The iterative scheme outlined here is based on that elaborated by
Hradil {\it et al.} for discrete-variable states \cite{Rehacek01,
Hradil97,Hradil99,Furasek01} . Here we give its brief overview.
Consider a large set of von Neumann measurements, each one
projecting the state of the system onto an eigenstate of a
measurement apparatus $\ket{y_j}$. The set of all possible
outcomes $\{\ket{y_j}\}$ can be associated with either one or
several measurement bases. Let $f_j$ be the frequency of
occurrences for each outcome. Then, with the system being in the
quantum state $\hat{\rho}$, the likelihood of a particular data
set $\{f_j\}$ is
\begin{equation} \label{Lgen}
\mathcal{L}(\hat\rho)=\prod_j{\rm pr}_j^{f_j},
\end{equation}
where ${\rm pr}_j=
\bra{y_j}\hat{\rho}\ket{y_j}=\Tr[\hat\Pi_j\hat{\rho}]$ is the
probability of the outcome $\ket{y_j}$ and
$\hat\Pi_j=\ket{y_j}\bra{y_j}$ denotes the projection operator.

To find the ensemble $\hat\rho$ which maximizes the likelihood
(\ref{Lgen}), one introduces the operator
\begin{equation} \label{Rgen}
\hat R(\hat{\rho})=\sum_j\frac{f_j}{{\rm pr}_j}\hat\Pi_j
\end{equation}
and notices that for the ensemble $\rho_0$ that is most likely to
produce the observed data set, $f_j\propto {\rm pr}_j$.
Furthermore, since the $\sum_j \hat\Pi_j\propto \hat 1$, we find
$\hat R (\hat{\rho}_0) \propto \hat 1$ and thus \begin{equation}
\label{Rrho}\hat R(\hat{\rho}_0)\hat\rho_0=\hat\rho_0\hat
R(\hat{\rho}_0)\propto\rho_0
\end{equation} as well as \begin{equation}  \label{RrhoR} \hat
R(\hat{\rho}_0)\hat\rho_0\hat R(\hat{\rho}_0)\propto\rho_0.
\end{equation}

The last relation forms the basis for the iterative algorithm. We
choose some initial denstity matrix as, e.g.,
$\hat\rho^{(0)}=\mathcal{N}[\hat 1]$, and apply repetitive
iterations
\begin{equation}\label{iterhomo}
\hat\rho^{(k+1)}=\mathcal{N}\left[\hat
R(\hat\rho^{(k)})\hat\rho^{(k)}\hat R(\hat\rho^{(k)})\right],
\end{equation}
where $\mathcal{N}$ denotes normalization to a unitary trace. Each
step will monotonically increase the likelihood associated with
the current density matrix estimate while the latter will
asymptotically approach the maximum-likelihood ensemble
$\hat\rho_0$\footnote[7]{We base our iteration scheme on Eq.
(\ref{RrhoR}) rather than (\ref{Rrho}) in order to ensure the
positivity of the density matrix at each step.}. This iterative
scheme can be viewed as a special case of the classical
expectation-maximization algorithm \cite{YardiLee}.

%Indeed, a homodyne measurement run yields a set of statistical
%data, whose likelihood with respect to a particular quantum
%ensemble can be evaluated easily. The goal of this section is to
%describe the way of finding the hypothetical ensemble which
%maximizes the probability of acquiring a particular data set. In
%writing this section we have assumed the reader to be familiar
%with the basics of quantum homodyne tomography; otherwise we
%recommend to refer to Chapter \ref{Raymer} of this book for a
%detailed review.

Now we turn to the main subject of the paper and consider a
homodyne tomography experiment performed on an optical mode
prepared in some quantum state $\hat\rho$. In an experimental run
one measures the field quadrature at various phases of the local
oscillator. Each measurement is associated with the observable
$\hat X_\theta = \hat X \cos\theta + \hat P \sin \theta$, where
$\hat X$ and $\hat P$ are the canonical position and momentum
operators and $\theta$ is the local oscillator phase.

For a given phase $\theta$, the probability to detect a particular
quadrature value $x$ is proportional to
\begin{equation}\label{pr}
{\rm pr}_\theta(x)={\rm Tr} [\hat\Pi(\theta,x)\hat\rho],
\end{equation} where $
\hat\Pi(\theta,x)=\ketbra{\theta,x}{\theta,x}$ is the projector
onto this quadrature eigenstate. In the Fock (photon number state)
basis, the projection operator is expressed as
\begin{equation}\label{projFock}
\Pi_{mn}(\theta,x)=\bra{m}\hat\Pi(\theta,x)\ket{n}=\braket{m}{\theta,x}\braket{\theta,x}{n},
\end{equation} where the overlap between the number and
quadrature eigenstates is given by the well known stationary
solution of the Schr\"odinger equation for a particle in a
harmonic potential:
\begin{equation}
\braket{n}{\theta,x} = e^{in\theta}
\left(\frac{2}{\pi}\right)^{1/4}\frac{H_n(\sqrt{2}x)}{\sqrt{2^n
n!}} \ \exp(-x^2),
\end{equation}
with $H_n$ denoting the Hermite polynomials\footnote[7]{
Normalization $[\hat x,\hat p]=i/2$ is used. The additional phase
factor $e^{im\theta}$ originates from the properties of the
phase-space rotation operator \cite{leon} $\hat
U(\theta)=e^{-i\theta \hat n}$. From $\hat U^\dagger(\theta)\hat
a\hat U (\theta)=\hat a e^{-i\theta}$ we find for the quadrature
operator $\hat U^\dagger(\theta)\hat X\hat U (\theta)=\hat
X_\theta$ and for its eigenstate $\ket{\theta,x}=\hat
U^\dagger\ket{0,x}$. From the first and last relations above, we
obtain $\braket{m}{\theta,x}=e^{i\theta m}\braket{m}{0,x}$. The
quantity $\braket{m}{0,x}$ is the energy eigenwavefunction of a
harmonic oscillator.}.

Because a homodyne measurement generates a continuous value, one
cannot apply the iterative scheme (\ref{iterhomo}) directly to the
experimental data. One way to deal with this difficulty is to
discretize the data by binning it up according to $\theta$ and $x$
and counting the number of events $f_{\theta,x}$ belonging to each
bin. In this way, a number of histograms, which represent the
marginal distributions of the desired ensemble's Wigner function,
can be constructed. They can then be used to implement the above
reconstruction procedure.

However, discretization of continuous experimental data will
inevitably lead to a loss of precision. To lower this loss, one
needs to reduce the size of a single bin and increase the number
of bins. In the limiting case of infinitely small bins,
$f_{\theta,x}$ takes on the values of either 0 or 1, so the
likelihood of a data set $\{(\theta_i,x_i)\}$ is given by
\begin{equation} \label{Lhomo}
\ln \mathcal{L} = \sum_i \ln {\rm pr}_{\theta_i}(x_i),
\end{equation}
and the iteration operator (\ref{Rgen}) becomes
\begin{equation}\label{Rhomo}
\hat R(\hat\rho) = \sum_i\frac{\hat\Pi(\theta_i,x_i)}{{\rm
pr}_{\theta_i}(x_i)},
\end{equation}
where $i=1\dots N$ enumerates individual measurements. The
iterative scheme (\ref{iterhomo}) can now be applied to find the
ensemble which maximizes the likelihood (\ref{Lhomo}).

In practice, the iteration algorithm is executed with the density
matrix in the photon number (Fock) representation. Since the
Hilbert space of optical states is of infinite dimension, the
implementation of the algorithm requires its truncation so the
Fock terms above a certain threshold are excluded from the
analysis. This assumption conforms to many practical experimental
situations in which the intensities of fields involved are {\it a
priori} limited.

%We note that the scheme above does not involve evaluating marginal
%distributions, i.e. histograms of the quadrature data
%$pr_\theta(x)$ associated with particular phases. Elimination of
%this intermediate step from the reconstruction scheme allows one
%to avoid approximating the phase and quadrature values and thus
%further enhance the accuracy of the method.

\begin{figure}[tbp]
\begin{center}
\includegraphics[width= 0.9 \columnwidth]{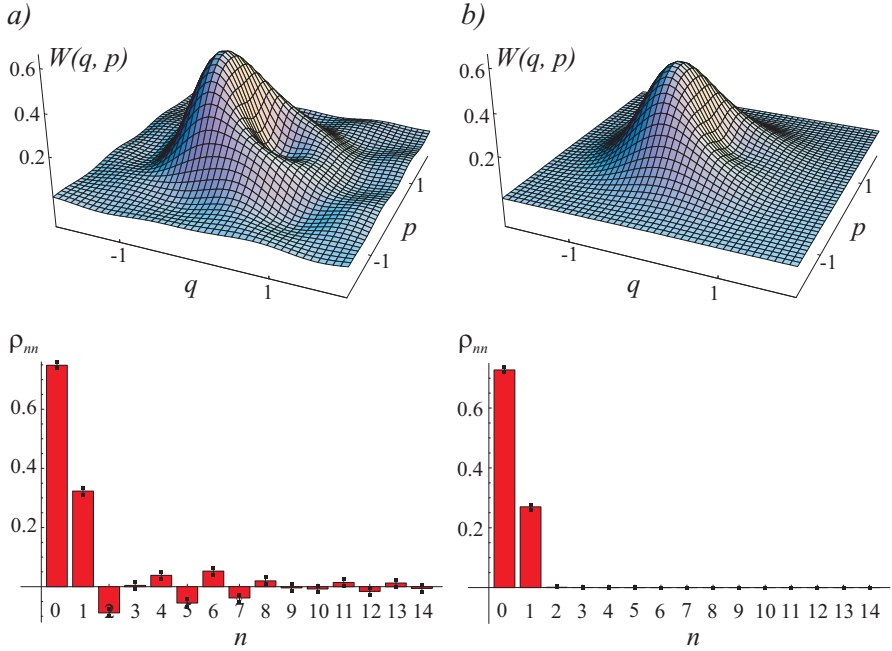}
\caption{\label{CompFig} Estimation of an optical ensemble from a
set of 14152 experimental homodyne measurements \cite{catalysis}
by means of the inverse Radon transformation (a) and the
likelihood maximization algorithm (b). The Wigner function and the
diagonal elements of the reconstructed density matrix are shown.
The inverse Radon transformation in (a) was performed by means of
the filtered back-projection algorithm with the cutoff frequency
of 6.3. The statistical uncertainties in (b) were determined by
means of a Monte-Carlo simulation (see text).}
\end{center}
\end{figure}

It is instructive to compare the maximum likelihood quantum state
estimation with the traditional methods employed in homodyne
tomography: reconstruction of the Wigner function by means of the
inverse Radon transformation \cite{leon} and evaluation of the
density matrix using quantum state sampling
\cite{patterns1,patterns2}. Fig. \ref{CompFig} shows application
of these two techniques to the experimental data from Ref.
\cite{catalysis}. The data set consists of 14152 quadrature
samples of an ensemble approximating a coherent superposition of
the single-photon and vacuum states.

The reconstruction shown in the figure reveals the advantages of
the MaxLik technique in comparison with the standard algorithm.
First, the finite amount and a discrete character of the data
available leads necessarily  to statistical noise which prevents
one from extracting complete information about a quantum state of
infinite dimension. To deal with this issue, both techniques apply
certain assumptions on the ensemble to be reconstructed. While the
MaxLik algorithm truncates the Fock space, the filtered
back-projection imposes low pass filtering onto the Fourier image
of the Wigner function\footnote[7]{The pattern function
reconstruction of the density matrix is free from this drawback as
it does not involve spectral filtering and relies on truncating
the Fock space instead.}, i.e. assumes the ensemble to possess a
certain amount of ``classicality" \cite{Vogel}. The latter
assumption is dictated by mathematical convenience and is much
less physically founded than the former. The ripples visible in
the Wigner function reconstruction in Fig. \ref{CompFig}(a) are a
direct consequence of statistical noise and are associated with
unphysical high number terms in the density matrix. Such ripples
are typical of the inverse Radon transformation \cite{iRt1,iRt2}.

Second, the back-projection algorithm does not impose any {\it a
priori} restrictions on the reconstructed ensemble. This may lead
to unphysical features in the latter, such as negative diagonal
elements of the density matrix in Fig. \ref{CompFig}(a). The
MaxLik technique, on the other hand, allows to incorporate the
positivity and unity-trace constraints into the reconstruction
procedure, thus always yielding a physically plausible ensemble
\cite{B1,B2}.

A third important advantage of the MaxLik technique is the
possibility to incorporate the detector inefficiences. In a
practical experiment, the photodiodes in the homodyne detector are
not 100\% efficient, i.e. they do not transform every incident
photon into a photoelectron. This leads to a distortion of the
quadrature noise behavior which needs to be adjusted for in the
reconstructed ensemble.

A common model for a homodyne detector of non-unitary efficiency
$\eta$ is a perfect detector preceded by a fictitious beam
splitter of transmission $\eta$. The reflected mode is lost so the
incident density matrix undergoes, in transmission through the
imaginary beam splitter, a so-called generalized Bernoulli
transformation \cite{leon}:
\begin{equation}
\bra{m}\hat\rho_\eta\ket{n}=\sum_{k=0}^\infty
B_{m+k,m}(\eta)B_{n+k,n}(\eta)\bra{m+k}\hat\rho_0\ket{n+k},
\end{equation}
where $\hat\rho_0$ and $\hat\rho_\eta$ are the density matrices of
the original and transmitted ensembles, respectively, and
$B_{n+k,n}=\sqrt{\left({n+k}\atop{n}\right)\eta^n(1-\eta)^k}$.
Under these circumstances the probability (\ref{pr}) of detecting
a quadrature value $x$ becomes
\begin{eqnarray}
\fl {\rm
pr}_\theta^\eta(x)&=&\bra{\theta,x}\hat\rho_\eta\ket{\theta,x}\\\fl&=&
\sum_{m,n=0}^\infty \sum_{k=0}^\infty
B_{m+k,m}(\eta)B_{n+k,n}(\eta)\braket{n}{\theta,x}\braket{\theta,x}{m}\bra{m+k}\hat\rho_0\ket{n+k},\nonumber
\end{eqnarray}
so the projection operator $\hat\Pi(\theta,x)$ becomes replaced by
a POVM element given by
\begin{eqnarray}
\fl \hat E_\eta(\theta,x)= \sum_{m,n,k}
B_{m+k,m}(\eta)B_{n+k,n}(\eta)\braket{n}{\theta,x}\braket{\theta,x}{m}\ket{n+k}\bra{m+k}.
\end{eqnarray}
Aware of the homodyne detector efficiency $\eta$, one runs the
iterative algorithm (\ref{iterhomo}) and reconstructs the original
density matrix $\hat\rho_0$ \cite{B5}.

Theoretically, it is also possible to correct for the detector
inefficiencies by applying the inverted Bernoulli transformation
\emph{after} an efficiency-uncorrected density matrix has been
reconstructed \cite{InvBernoulli}. However, this may give rise to
unphysically large density matrix elements associated with high
photon numbers. With the inefficiency correction incorporated, as
described above, into the reconstruction procedure, such terms do
not arise \cite{B2}.

%The performance of this technique is illustrated by Fig.
%\ref{BanaszeksFig} showing the reconstructed photon statistics of
%a coherent state measured with an inefficient detector. The
%estimated photon number distribution represents correctly the
%original state, in contrast to that obtained via the quantum state
%sampling followed by the inverted Bernoulli transformation
%\cite{InvBernoulli}.

It is interesting to discuss the MaxLik reconsruction of the
density matrix in comparison with the point-by-point
reconstruction of the Wigner function as proposed in \cite{B3}. To
determine the value of the Wigner function at a specific point
$(p,q)$ in the phase space, Ref. \cite{B3} proposes to apply a
phase-dependent shift to the experimental data which corresponds
to a displacement of the point $(p,q)$ to the phase space origin.
Then one reconstructs a phase-averaged ensemble according to Refs.
\cite{B1,B2}, and calculates the value of the Wigner function at
the origin of the displaced phase space, which is equal to that at
the  desired location in the undisplaced phase space.

This scheme may appear more involved than the one proposed here,
as one needs to run a separate iteration series for every point in
which the Wigner function is to be calculated. However, due to a
smaller number of parameters and a simplified iteration step, each
iteration takes less time and the series converges faster. The
choice of a particular scheme depends on a specific task and on
the chosen truncation threshold in the Fock space. It is important
to note that the scheme \cite{B3} does not impose any \emph{a
priori} restrictions onto the ensemble being reconstructed, and
therefore the latter is not guaranteed to be physically
meaningful.

%For such an ensemble, the probability (\ref{pr}) of detecting a
%particular quadrature value is also independent from the local
%oscillator phase:
%\begin{equation}
%{\rm pr}(x)=\sum_{n=0}^\infty\rho_{nn}\braket{n}{x}^2
%\end{equation}
%Using the above relation we

Finally, we discuss statistical uncertainties of the reconstructed
density matrix. In generic MaxLik algorithms, they are typically
estimated as an inverse of the Fisher information matrix
\cite{Rao,Cramer} $G=\partial^2\mathcal{L}(\vec t)/\partial\vec
t\partial\vec t'$, where $\vec t$ denotes a set of independent
parameters with respect to which the likelihood is evaluated.
Because the density matrix elements are not fully independent but
bound by the positivity and unity trace constraints, one expresses
the density matrix as a product $\hat\rho=\hat T^\dagger\hat
T/\Tr[\hat T^\dagger\hat T]$. Now the constraints for $\hat\rho$
are satisfied for any random $\hat T$ and one can regard the
elements of the latter as free parameters in evaluating the Fisher
information \cite{B5,Usami}.

%, and the Fisher information matrix can be written as
%\begin{equation} \hat F(\hat\rho_{ML})=\frac{partial^2}

%However, this method is not fully applicable to MaxLik schemes
%involving \emph{a priori} constraints, such as the density matrix
%positivity constraint relevant here. If one of the parameters lies
%near the boundary of the positivity region, the statistical error
%will be governed by the first rather than the second derivative of
%the likelihood \cite{B5}, and the actual uncertainty will be
%smaller than the one estimated.

A sensible alternative is offered by a clumsy, yet simple and
robust technique of {\it simulating} the quadrature data that
would be associated with the estimated ensemble $\hat\rho_{ML}$ if
it were the true state. One generates a large number of random
sets of homodyne data according to Eq. (\ref{pr}), then applies
the MaxLik reconstruction scheme to each set and obtains a series
of density matrices $\hat\rho'_k$ each of which approximates the
original matrix $\hat\rho_{ML}$. The average difference $\langle|
\hat\rho_{ML} - \hat\rho'_k|\rangle_k$ evaluates the statistical
uncertainty associated with the reconstructed density matrix.

\ack{This work was supported by the Deutsche
Forschungsgemeinschaft and the Optik-Zentrum Konstanz. I am
grateful to Z Hradil, J \v{R}eh\'{a}\v{c}ek, Hradil, M Je\v{z}ek,
G M D'Ariano and P. Lo Presti for stimulating discussions.}

\end{document}